\documentclass{ws-mplb}
\usepackage{amsfonts,amssymb,amsmath}
\usepackage[super,sort,compress]{cite} 
\usepackage[pdftex]{graphicx,color}

\begin{document}

\markboth{F. Okubo and H. Katsuragi}{Force chain structure in a rod-withdrawn granular layer}

%
\catchline{}{}{}{}{}
%

\title{Force chain structure in a rod-withdrawn granular layer}

\author{F. Okubo}

\address{Department of Earth and Environmental Sciences, Nagoya University, Furocho, Chikusa, Nagoya 464-8601, Japan}

\author{H. Katsuragi}

\address{
Department of Earth and Space Science, Osaka University, 1-1 Machikaneyama, Toyonaka 560-0043, Japan\\
}

\maketitle

\begin{history}
\received{(Day Month Year)}
\revised{(Day Month Year)}
\end{history}

\begin{abstract}
When a rod is vertically withdrawn from a granular layer, oblique force chains can be developed by effective shearing. In this study, the force-chain structure in a rod-withdrawn granular layer was experimentally investigated using a photoelastic technique. The rod is vertically withdrawn from a two-dimensional granular layer consisting of bidisperse photoelastic disks. During the withdrawal, the development process of force chains is visualized by the photoelastic effect. By systematic analysis of photoelastic images, force chain structures newly developed by the rod withdrawing are identified and analyzed. In particular, the relation between the rod-withdrawing force $F_\mathrm{w}$, total force-chains force $F_\mathrm{t}$, and their average orientation $\theta$ are discussed. We find that the oblique force chains are newly developed by withdrawing. The force-chain angle $\theta$ is almost constant (approximately $20^{\circ}$ from the horizontal), and the total force $F_\mathrm{t}$ gradually increases by the withdrawal. In addition, $F_\mathrm{t}\sin\theta$ shows a clear correlation with $F_\mathrm{w}$.
\end{abstract}


\section{Introduction}

Granular assembly is well known as a simple but peculiar process that exhibits various intriguing phenomena. For instance, granular matter shows a phase-transition-like behavior called a jamming transition~(e.g., Ref.~\citen{Behringer:2014}). Basically, the jamming transition is similar to the liquid-solid transition without ordering. In addition to the usual jamming transition, a shear-induced jamming transition was also found~\cite{Bi:2011}. Although shearing usually weakens a solidified structure, shear jamming can enhance the solidification of macroscopic granular matter. In sheared granular matter, characteristic force chain structures are formed~\cite{Majmudar:2005,Ren:2013}, and their relation to shear jamming has been discussed~\cite{Wang:2018}. To visualize force-chain structures, photoelastic disks have been frequently used since Oda developed the method almost a half century ago~\cite{Oda:1974}.   

Another salient feature observed in granular matter is complex friction. The granular friction coefficient $\mu$ becomes a function of a dimensionless number called the inertial number $I=\dot{\gamma}d/\sqrt{p/\rho}$, where $\dot{\gamma}$, $d$, $p$, and $\rho$ are the shear strain rate, diameter of grains, confining pressure, and density of grains, respectively. Such characterization of the granular friction is different from that of rock friction with a rate-state-dependent law~\cite{Marone1998}. Various types of granular friction have been modeled with the variable $I$~\cite{GDRMiDi:2004,daCruz2005,Jop2006,Pouliquen2006}. This so-called $\mu-I$ rheology of granular friction is one of the most popular basis for considering complex granular behaviors. For an advanced study, for instance, velocity weakening of granular friction was observed in a small $I$ regime~\cite{Kuwano2013}.  The friction coefficient $\mu$ is defined by the ratio between the normal force $F_\mathrm{norm}$ and tangential force $F_\mathrm{tan}$, $\mu = F_\mathrm{tan}/F_\mathrm{norm}$. Namely, shearing by a tangential force is necessary to characterize the friction. As mentioned above, sheared granular matter can show a certain kind of solidification: shear jamming. This phenomenon can affect its frictional behavior owing to the shearing itself. 

Furuta et al. conducted a simple experiment to study the relationship between granular shearing and friction. In the experiment, a rod was vertically withdrawn from a three-dimensional (3D) glass-bead layer~\cite{Furuta:2017}. In their subsequent study~\cite{Furuta:2019}, a similar rod-withdrawing experiment was performed by precisely controlling the packing fraction $\phi$ of the granular layer using air fluidization and mechanical vibration~\cite{Furuta:2019}. In Ref.~\citen{Furuta:2019}, the divergent behavior of the withdrawing force $F_\mathrm{w}$ was observed as $\phi$ approached its critical value. Based on this experimental result, Furuta et al. considered the shear-induced development of an effectively solidified zone supported by force chains around the withdrawn rod. This idea is consistent with the experimental data and is supported by numerical simulations ~\cite{Furuta:2019}. However, no direct measurement of the corresponding force-chain development has been experimentally carried out. 

Therefore, in this study, we performed an experiment in which a rod is vertically withdrawn from a granular layer. To characterize the force-chain development, we employed the photoelastic technique. Although the system is restricted to a two-dimensional (2D) one, the use of photoelasticity is advantageous for visualizing the state of granular internal forces. From the experimental results, we qualitatively evaluate the validity of the force-chain development proposed in Ref.~\citen{Furuta:2019}.

\section{Experiment}
\subsection{Experimental setup}
\begin{figure}[th]
\begin{center}
\includegraphics[width=.6\linewidth]{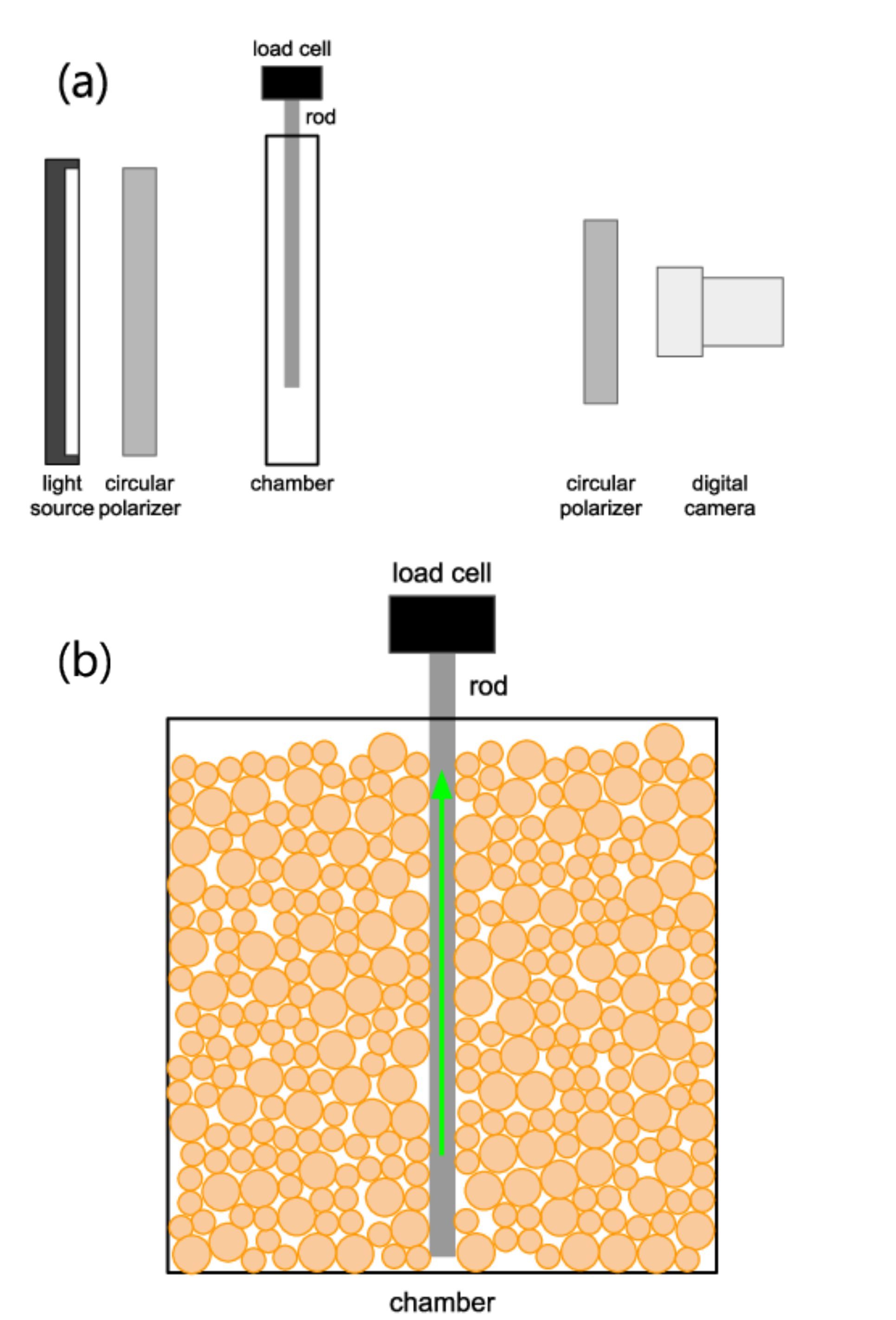}
\end{center}
\vspace*{8pt}
\caption{Schematic images of experimental apparatus. (a) Sideview and (b) frontview images show optical setup and configuration of experimental cell and photoelastic disks, respectively. \label{fig:apparatus}}
\end{figure}

The experimental setup built to observe the granular internal force structure is shown in Fig.~\ref{fig:apparatus}. Figure~\ref{fig:apparatus}(a) shows a schematic of the system in which the optical setup is presented. The front view of the experimental system is schematically shown in Fig.~\ref{fig:apparatus}(b). A 2D cell with inner dimensions of $0.3 \times 0.3 \times 0.011$~m$^3$ made of acrylic plates was vertically held. A rod (diameter $D=6$~mm) was vertically placed at the center of the cell and connected to the load cell of a universal testing machine (Shimadzu AG-X). Then, the cell was filled with a bidisperse set of photoelastic disks (Vishay Micromeasurements, PSM-4) with diameters of $D_L=15$~mm or $D_S=10$~mm. The thickness of these disks was $10$~mm. To fill the cell, 200 large and 400 small disks were used. The filling was manually performed to prepare a loosely packed random initial state. To control the initial packing fraction $\phi_0$, mechanical tapping was manually added after the filling procedure. The number of taps was carefully controlled to achieve the desired $\phi_0$ value. For a withdrawn rod, we used three different surface types: threaded, roughened by filing, and covered by a frictional sheet. However, all experimental results were insensitive to the surface of the rod, as demonstrated later. The experimental cell was illuminated by a uniform light source from the back, and the transmitted light was acquired by a digital still camera (Nikon N7100). Images of $6000 \times 4000$ pixels in size could be captured by this system with a spatial resolution of $76.1$~${\mu}$m/pixel. 

An example photo of the experimental cell with photoelastic disks is shown in Fig.~\ref{fig:raw_data}(a). Using this bright-field image, we measured the position of the disks and the initial packing fraction $\phi_0$. Sandwiching the cell with two circular polarizers placed at right angles, a corresponding dark-field image shown in Fig.~\ref{fig:raw_data}(b) can be obtained. To reduce extraneous noise, images were captured in a dark room. Then, owing to the birefringence induced by the photoelastic effect, the applied force at each contact point was clearly visualized, as shown in Fig.~\ref{fig:raw_data}(b). In the image, one can see the very inhomogeneous internal force distribution called the force chain structure. 

\subsection{Experimental procedure}
\begin{figure}[th]
\begin{center}
\includegraphics[width=.6\linewidth]{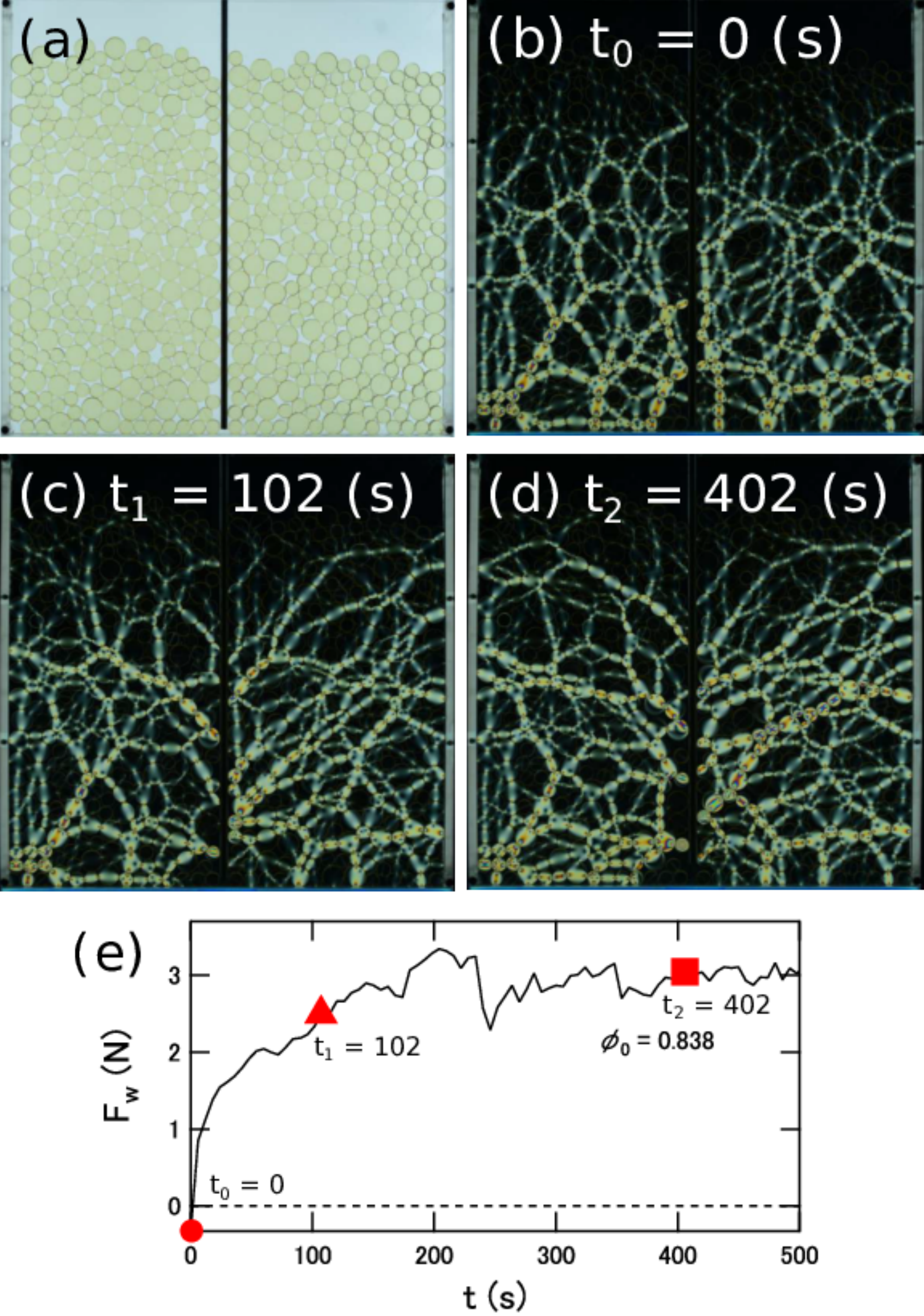}
\end{center}
\vspace*{8pt}
\caption{Raw data of (a) initial bright-field image, (b-d) dark-field images, and (e) rod withdrawing force data $F_\mathrm{w}(t)$. Panels (b), (c), and (d) correspond to $t=0$ (initial state), $t=102$~s, and $t=402$~s, respectively, where $t$ represents elapsed time from beginning of withdrawal. Initial packing fraction in this experimental run was $\phi_0=0.838$. \label{fig:raw_data}}
\end{figure}

Before withdrawing the rod, the initial packing fraction $\phi_0$ was measured from the bright-field image. To reduce the boundary effect, $\phi_0$ was computed in the central region of the cell. Specifically, the boundary region of thickness $D_L$ from the rod, walls, or top surface was removed from the analysis. $\phi_0$ was simply measured by the ratio of the area occupied by disks and the area of the void. In this study, $\phi_0$ was varied in the range of $0.814$--$0.842$. Then, the rod was slowly withdrawn from the cell at a constant rate of $v = 0.025$~mm/s. The measured withdrawing force $F_\mathrm{w}$ as a function of time $t$ is shown in Fig.~\ref{fig:raw_data}(e), where $t=0$ corresponds to the beginning instance of the withdrawing. A sampling rate of $F_\mathrm{w}$ was 100 Samples/s and $F_\mathrm{w}$ was measured during $0\leq t \leq 500$~s. Note that the withdrawing distance at $t=500$~s corresponds to $1.25D_s$. Namely, we focused on the very early stage of the rod withdrawing. As seen in Fig.~\ref{fig:raw_data}(e), $F_\mathrm{w}$ shows an initial increase followed by an almost steady $F_\mathrm{w}$ regime. This behavior is consistent with the 3D case~\cite{Furuta:2017,Furuta:2019}.

Regarding image acquisition, although the complete simultaneous data acquisition of both bright- and dark-field images is impossible, we alternately acquired bright- and dark-field images every 3~s. Because the withdrawing rate is very slow, practically simultaneous data acquisition can be achieved by this protocol. Using both bright- and dark-field images, we can measure the force applied to each photoelastic disk. Examples of dark-field images taken at $t=102$~s and $t=402$~s are shown in Figs.~\ref{fig:raw_data}(c) and (d), respectively.

\subsection{Calibration of photoelastic images}
To quantitatively estimate the contact force exerted on the photoelastic disks, calibration is necessary. We employed a calibration method that uses the squared intensity gradient per disk $\langle G^2 \rangle=\langle |\nabla I|_\mathrm{disk}^2 \rangle$, where $I_\mathrm{pe}$ is the photoelastic intensity, and $|\nabla I_\mathrm{pe}|_\mathrm{disk}$ denotes the absolute intensity gradient per disk~\cite{Howell:1999}. Using the calibration protocol mentioned in Refs.~\citen{Iikawa:2015,Iikawa:2018}, we obtained the relation between the contact force per disk $F_d$ and $\langle G \rangle$ as $F_d = 1.16 \times 10^{-7} \langle G^2 \rangle^{4.7}$ for large disks, and $F_d = 1.31 \times 10^{-10} \langle G^2 \rangle^{6.0}$ for small disks under the current experimental conditions.

\section{Results and analyses}
In the raw data shown in Fig.~\ref{fig:raw_data}, the force chain structures are visible even in the initial state $t=0$~(Fig.~\ref{fig:raw_data}(b), indicating that the force chain is developed by the self-gravity of disks (hydrostatic force). Although we can observe the strengthening of the force chain structures in Figs.~\ref{fig:raw_data}(c) and (d), the initial self-gravity-based force chains camouflage the newly developed force chains. Although an increase in $F_\mathrm{w}$ can be clearly observed in Fig.~\ref{fig:raw_data}(e), the identification of the corresponding force-chain strengthening is not easily noticed in Figs.~\ref{fig:raw_data}(c) and (d).

\subsection{Force-chain structure analysis}
\begin{figure}[th]
\begin{center}
\includegraphics[width=.6\linewidth]{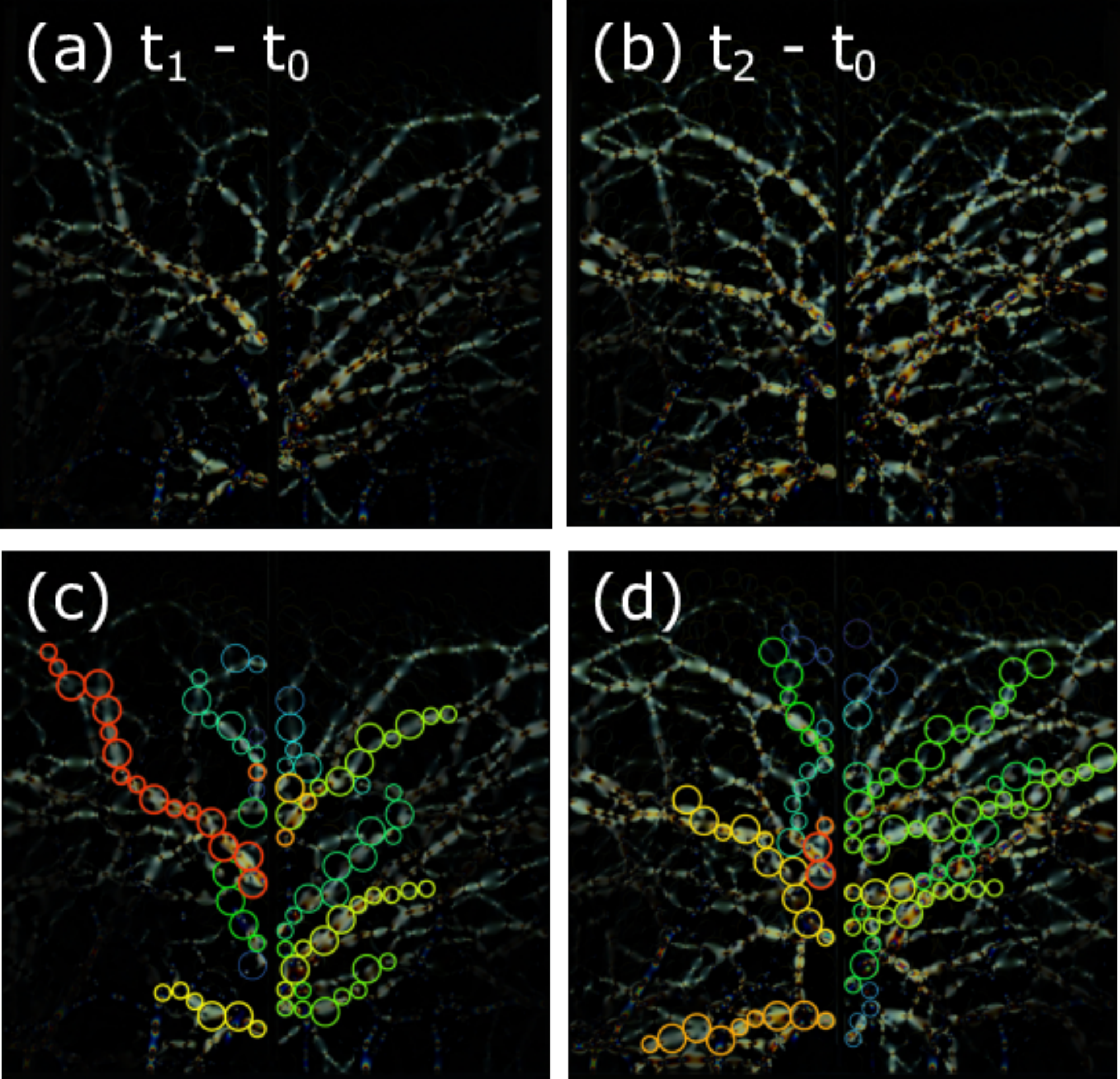}
\end{center}
\vspace*{8pt}
\caption{(a,b) Subtracted force-chain images and (c,d) identified force-chain structures. Subtracted images were computed from images shown in Fig.~\ref{fig:raw_data}. In panels (c) and (d), identified force chains are superimposed on subtracted images shown in panels (a) and (b), respectively. Identical color circles belong to same force chain. \label{fig:diff_data}}
\end{figure}

In order to remove the background force chains owing to self-gravity, we subtracted the initial dark-field image ($t=0$) from all other dark-field images. Examples of the subtracted images are shown in Figs.~\ref{fig:diff_data}(a) and (b). Figures~\ref{fig:diff_data}(a) and (b) correspond to the images in ``Fig.~\ref{fig:raw_data}(c)$-$Fig.~\ref{fig:raw_data}(b)'' and ``Fig.~\ref{fig:raw_data}(d)$-$Fig.~\ref{fig:raw_data}(b)'', respectively. In these images, newly developed force chains can be clearly identified. Note that the negative intensity values are shown in black (zero intensity) in Figs.~\ref{fig:diff_data}(a) and (b). Thus, the disappearing force chains are not visualized in this method. As seen in Figs.~\ref{fig:diff_data}(a) and (b), newly developed force chains obliquely extend from the rod toward the sidewalls. This force-chain structure is qualitatively consistent with the model proposed in Ref.~\citen{Furuta:2019}.

\begin{figure}[th]
\begin{center}
\includegraphics[width=.6\linewidth]{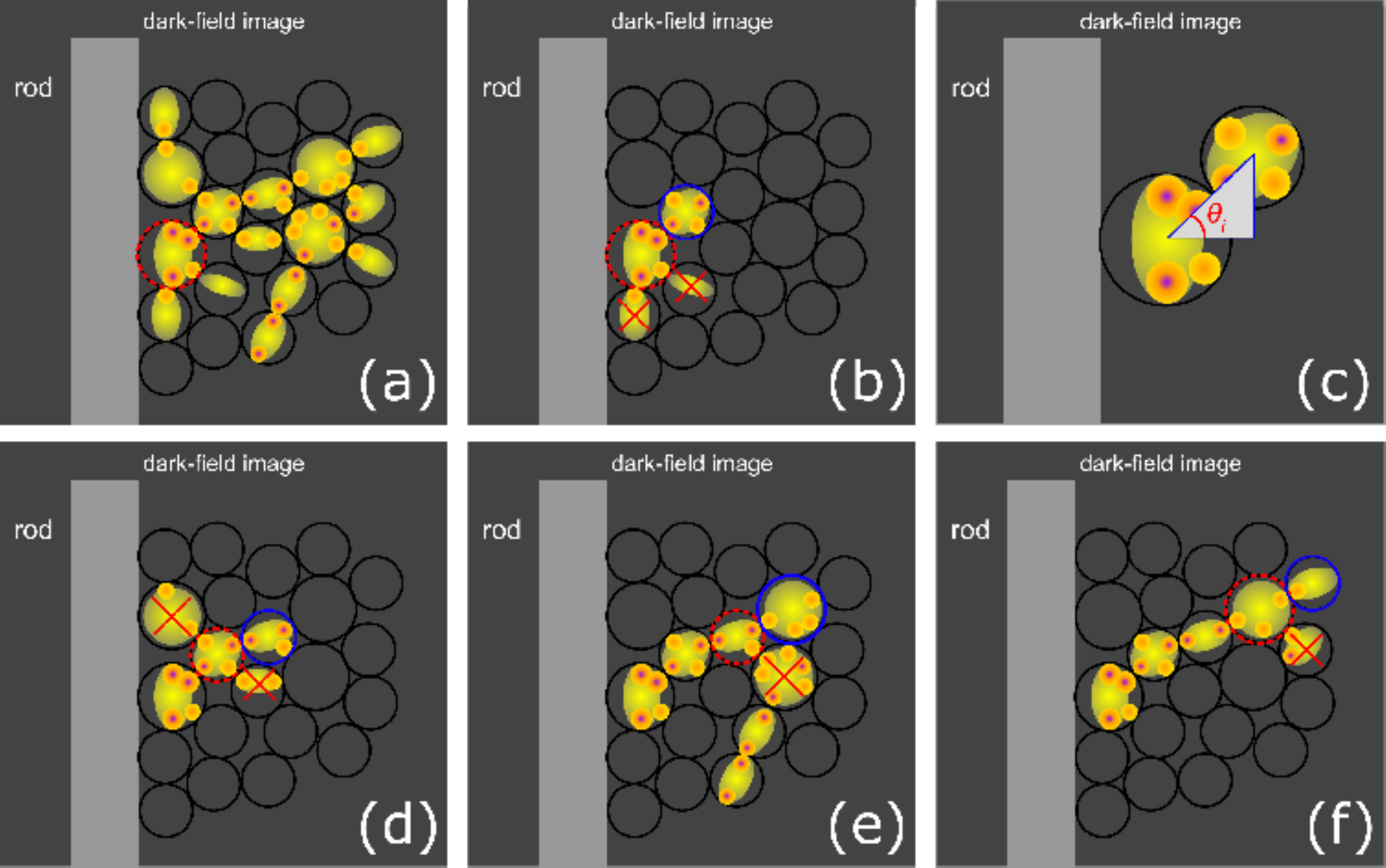}
\end{center}
\vspace*{8pt}
\caption{Schematic illustration of force-chain tracing algorithm. (a)~Starting point, (b)~strongest connection tracing, (c)~definition of connection angle, and (d-f) examples of connection tracing. Red dashed circles and blue circles are connected. \label{fig:FC_trace}}
\end{figure}

To further quantify the force chain structure, we must identify the force chains. For this purpose, we traced the force chains using the algorithm described below. Examples of force-chain tracing are schematically shown in Fig.~\ref{fig:FC_trace}. 

\begin{romanlist}[(ii)]
\item First, disks contacting with the rod were picked up as candidates for the starting points of force chains. Among the candidates, disks with contact force $F_\mathrm{d}$ greater than the threshold $F_\mathrm{th}(=7.5\times 10^{-2}$~N) were considered as actual starting points. For example, the red circular disk in Fig.~\ref{fig:FC_trace}(a) was selected as the starting point. Here, the starting point of the force chain is indexed by $i=0$, that is, $i$ denotes the generation of the disk in the force chain. 
\item Next, the skeleton structure of the force chains must be traced. To trace the 1st-generation disk ($i=1$), the strongest connection except for the vertically aligned disk was picked up, as shown in Fig.~\ref{fig:FC_trace}(b). Here, the angle of the connection $\theta_i$ is defined as shown in Fig.~\ref{fig:FC_trace}(c). Only the right region is shown in Fig.~\ref{fig:FC_trace}. The angle is mirror-symmetrically defined in the left region as well, that is, upward inclination from the horizon corresponds to the positive $\theta_i$, and $\theta_i$ ranges from $-90^{\circ} < \theta_i < 90^{\circ}$.  
\item Basically, the above process is iterated until the force chain reaches the boundary or all connecting disks satisfy $F_\mathrm{d}<F_\mathrm{th}$. However, here, we additionally assume the linear tendency of the force-chain skeleton structure. To identify the $n$th-generation connection, the average angle of the force chain $\theta_\mathrm{ave}=\sum_{i=1}^{n-1} \theta_i /(n-1)$ is introduced (for $n\geq 2$). To take into account the preference for linearity, two thresholds $\Delta \theta_\mathrm{th1}=30^{\circ}$ and $\Delta \theta_\mathrm{th1}=60^{\circ}$ are considered. The connections satisfying $\theta_n - \theta_\mathrm{ave} < \Delta \theta_\mathrm{th1}$ (and $F_\mathrm{d} \geq F_\mathrm{th}$) have first priority to be selected as an $n$th-generation disk. If there is no disk satisfying the above condition, the connections satisfying $\theta_n - \theta_\mathrm{ave} < \Delta \theta_\mathrm{th2}$ (and $F_\mathrm{d} \geq F_\mathrm{th}$) are inspected. Further, if there is no disk satisfying the above conditions, all connecting disks with $F_\mathrm{d} \geq F_\mathrm{th}$ are considered as candidates for the $n$th-generation disk. For example, connections are identified as shown by the blue circles in Fig.~\ref{fig:FC_trace}(d-f). In this algorithm, branching of the force chain is not permitted. However, each disk can belong to the force chains up to twice. In addition, the loop structure of the force chain detouring back to the rod is excluded. 
\end{romanlist}

The abovementioned algorithm was applied to the subtracted images to identify the newly developed force chains. In Figs.~\ref{fig:diff_data}(c) and (d), the identified force chains are shown as colored circles. The same-color circles belong to the same force chain. As seen in Fig.~\ref{fig:diff_data}, the principal backbone structure of the force chains can be appropriately identified by the abovementioned tracing algorithm.

\subsection{Temporal evolution of forces and angle}  
To quantitatively analyze the force chain structures, we computed two quantities from all force chains. One is the total force $F_\mathrm{t}$, which is defined by the summation of $F_\mathrm{d}$ composing the force chain. The other is the average angle $\theta$ of the force chain, which is computed by the average of $\theta_i$.

\begin{figure}[th]
\centerline{\includegraphics[width=7cm]{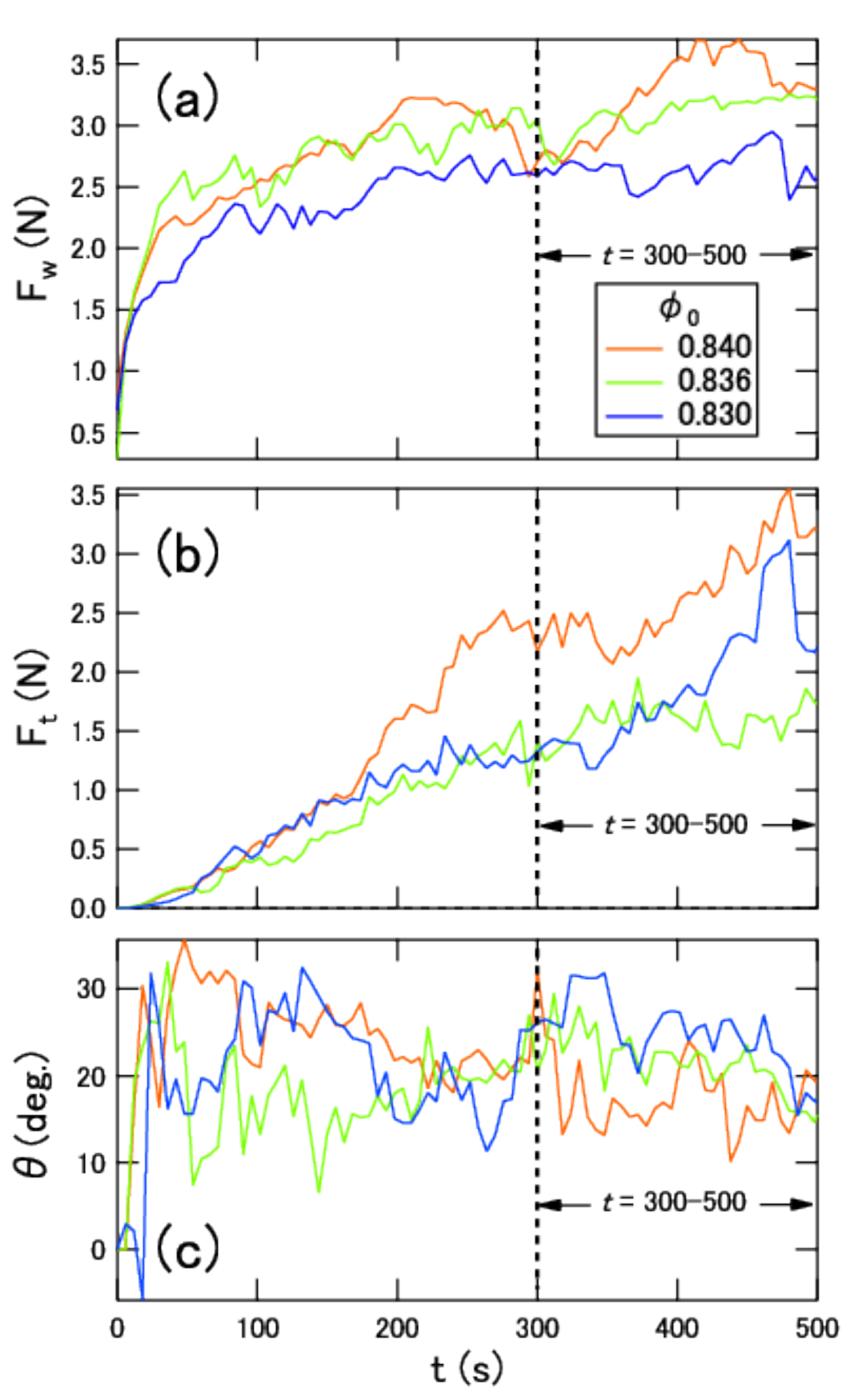}}
\begin{center}
\end{center}
\vspace*{8pt}
\caption{Temporal variations of (a) rod-withdrawing force $F_\mathrm{w}$, (b) total force-chains force $F_\mathrm{t}$, and (c) force-chains average angle $\theta$. Color indicates initial packing fraction $\phi_0$. \label{fig:t_variation}}
\end{figure}

Using the method defined so far, the temporal evolution of the force chain structure and its relation to the withdrawal force can be analyzed. In Fig.~\ref{fig:t_variation}, examples of temporal evolutions of $F_\mathrm{w}$, $F_\mathrm{t}$, and $\theta$ are shown. As can be seen, all of these quantities increase as the rod withdrawal proceeds. However, the qualitative behaviors of $F_\mathrm{w}(t)$, $F_\mathrm{t}(t)$, and $\theta(t)$ are different. $F_\mathrm{w}(t)$ gradually increases and approaches an asymptotically steady value (Fig.~\ref{fig:t_variation}(a). As seen in Fig.~\ref{fig:t_variation}(b), the increasing trend of $F_\mathrm{t}(t)$ is calmer than that of $F_\mathrm{w}(t)$ in the early stage. However, the increasing trend continues longer than $F_\mathrm{w}$. By contrast, $\theta(t)$ increases very rapidly and seems to fluctuate around the asymptotic value $\simeq 20^{\circ}$. Based on the law of action and reaction, we expect that the $F_\mathrm{w}$ behavior should be explained by a certain combination of the two effects of $F_\mathrm{t}$ and $\theta$.  

\section{Discussion}
\begin{figure}[th]
\centerline{\includegraphics[width=7cm]{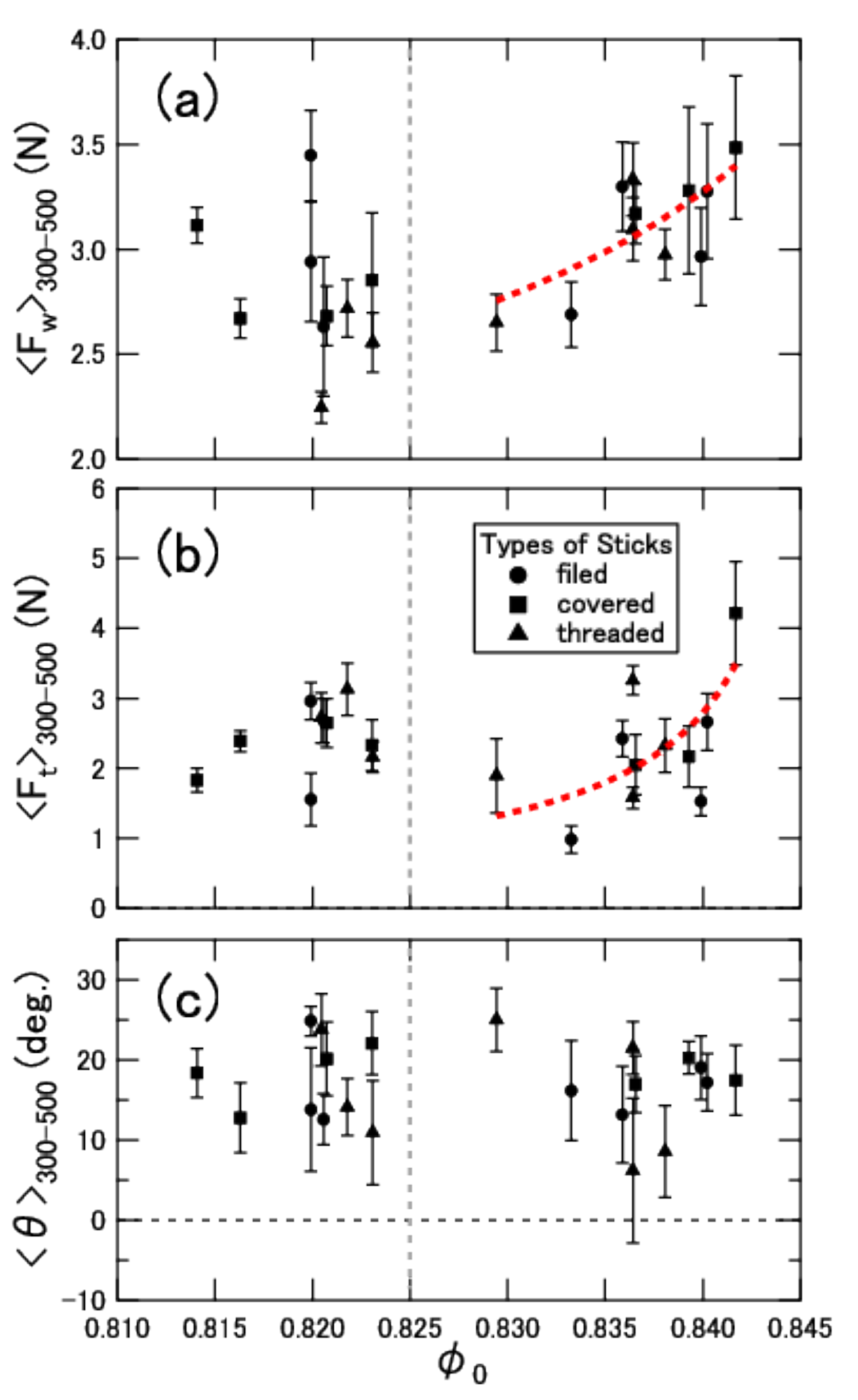}}
\begin{center}
\end{center}
\vspace*{8pt}
  \caption{Initial packing fraction $\phi_0$ dependences of (a) average rod-withdrawing force $\langle F_\mathrm{w} \rangle_{300\mbox{--}500}$, (b) average force-chains force $\langle F_\mathrm{t} \rangle_{300\mbox{--}500}$, and average force-chains angle $\langle \theta \rangle_{300\mbox{--}500}$. All quantities are averaged in approximately steady withdrawing regime, $300 \leq t \leq 500$~s. Dashed vertical lines distinguish small and large $\phi_0$ regimes. Dashed curves in (a) and (b) are guides to the eye. Symbols indicate surface conditions of rod as labeled in legend. \label{fig:phi0_dependence}}
\end{figure}

To characterize the shear-induced force-chain structure in the steady withdrawing regime, here, we focus on the relatively late stage, $300 \leq t \leq 500$~s. Although the initial transient behavior of force chain development is interesting, the steady behavior should be first analyzed to compare the result with previous studies~\cite{Furuta:2017,Furuta:2019}. In Ref.~\citen{Furuta:2019}, the $\phi_0$-dependent growth of force chains was assumed. To check the validity of this assumption, we compute the average quantities in the quasisteady state ($300 \geq t \geq 500$), $\langle F_\mathrm{w} \rangle_{300\mbox{--}500}$, $\langle F_\mathrm{t} \rangle_{300\mbox{--}500}$, and $\langle \theta \rangle_{300\mbox{--}500}$. These quantities are plotted as functions of $\phi_0$ in Fig.~\ref{fig:phi0_dependence}. Two distinct regimes can be seen in Fig.~\ref{fig:phi0_dependence}. In the relatively small $\phi_0 (<0.825)$ regime, the data scattering is significant, and it is difficult to see any systematic trend. In the relatively large $\phi_0 (>0.825)$ regime, on the other hand, positive correlation between forces ($\langle F_\mathrm{w} \rangle_{300\mbox{--}500}$ and $\langle F_\mathrm{t} \rangle_{300\mbox{--}500}$) and $\phi_0$ can be seen. However, the average force-chain angle $\langle \theta \rangle_{300\mbox{--}500}$ does not show any trend in the entire $\phi_0$ regime. Thus, we consider that the typical inclination angle of force chains ($\theta \simeq 20^{\circ}$) induced by the rod withdrawing is quite robust and independent of the initial conditions. By contrast, the force chain structure is strengthened by increasing $\phi_0$. As can be seen in Fig.~\ref{fig:phi0_dependence}, the surface conditions of the withdrawn rod do not affect the measured result.

Although the $\phi_0$-dependent increasing tendency of $F_\mathrm{w}$ (Fig.~\ref{fig:phi0_dependence}(a)) is consistent with the previous work~\cite{Furuta:2019}, its variation range is very narrow in the current experiment. The reason for this weak increase in $F_\mathrm{w}$ is probably the limited number of disks we used. In a previous work~\cite{Furuta:2019}, a 3D system with small (mainly less than 1~mm in diameter) glass beads was used. However, in this study, we used large disks (diameter greater than 10~mm) in a 2D system. Whereas we can visualize the structure of force chains owing to the photoelastic effect, the number of disks used in the experiment was limited in order to use this technique. 

In addition, we cannot observe the systematic behavior in the small $\phi_0$ regime, as shown in Fig.~\ref{fig:phi0_dependence}. Because we manually prepared the initial configuration of photoelastic disks, the initial conditions were significantly influenced by the fluctuation due to the strong protocol dependence of granular behavior. In general, granular behavior strongly depends on its preparation methods and shows large fluctuations~\cite{Jotaki:1977,Bertho:2003,Katsuragi:2016}. The large fluctuation observed in the small $\phi_0$ regime in Fig.~\ref{fig:phi0_dependence} could originate from the protocol-dependent fluctuation of granular matter. Therefore, in the following analysis, we concentrate on the behaviors in a relatively large $\phi_0(>0.825)$ regime. 

\begin{figure}[th]
\centerline{\includegraphics[width=7cm]{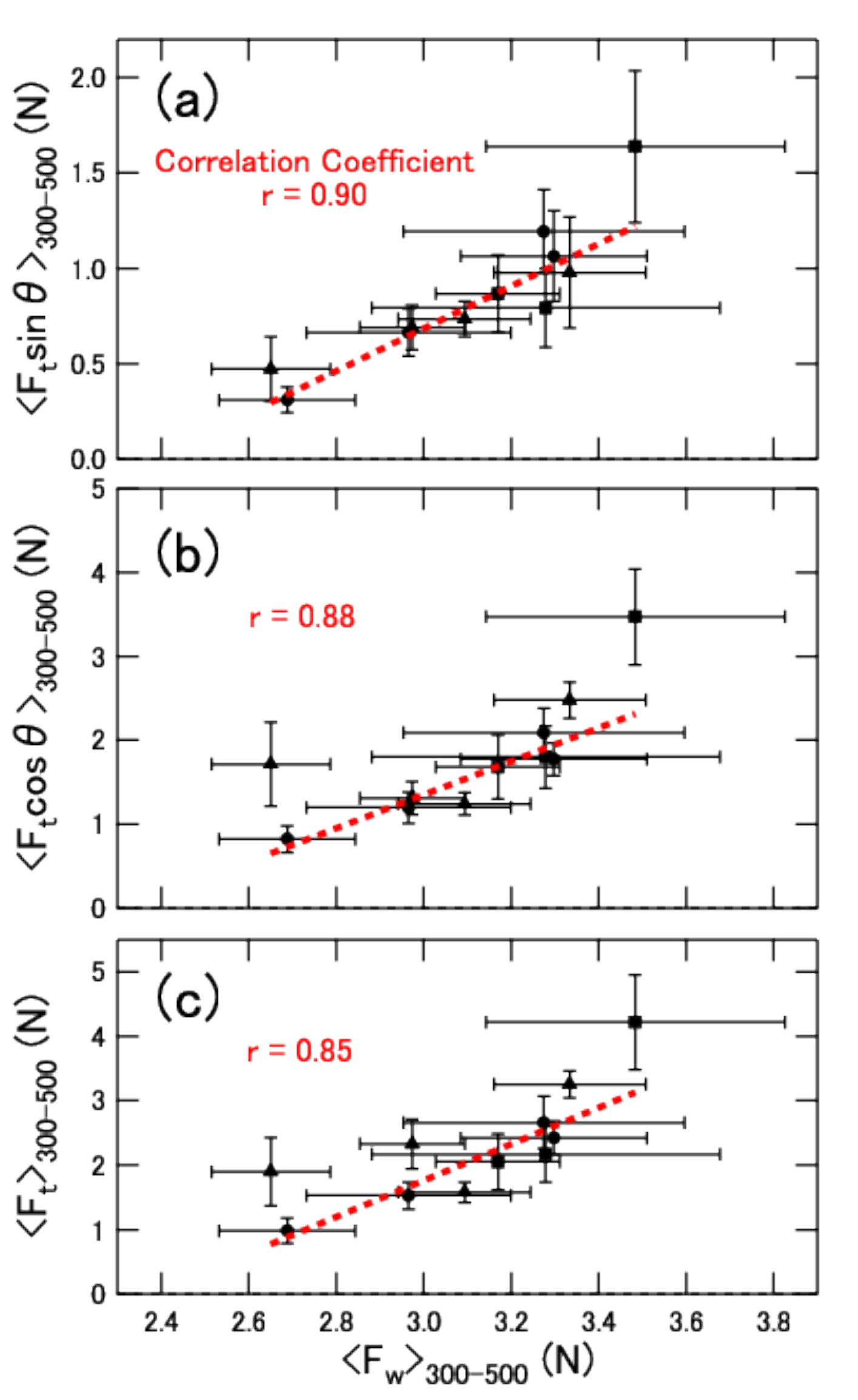}}
\begin{center}
\end{center}
\vspace*{8pt}
  \caption{Correlations between (a) shear component $\langle F_\mathrm{t}\sin\theta \rangle_{300\mbox{--}500}$, (b) normal component $\langle F_\mathrm{t}\cos\theta \rangle_{300\mbox{--}500}$, and (c) raw total force $\langle F_\mathrm{t} \rangle_{300\mbox{--}500}$ with rod-withdrawing force $\langle F_\mathrm{w} \rangle_{300\mbox{--}500}$ in steady withdrawal and large $\phi_0$ regime. Dashed lines indicate linear fitting. $r$ values denote correlation coefficients. \label{fig:correlations}}
\end{figure}

To consider the relation between $F_\mathrm{w}$, $F_\mathrm{t}$, and $\theta$, we plot $\langle F_\mathrm{t}\sin\theta \rangle_{300\mbox{--}500}$, $\langle F_\mathrm{t}\cos\theta \rangle_{300\mbox{--}500}$, and $\langle F_\mathrm{t} \rangle_{300\mbox{--}500}$ as functions of $\langle F_\mathrm{w}\rangle_{300\mbox{--}500}$. This is shown in Fig.~\ref{fig:correlations}. Note that only the data in the large $\phi_0(>0.825)$ regime are plotted in Fig.~\ref{fig:correlations} The factor $\sin\theta$ represents the shear component of $F_\mathrm{t}$, and $\cos\theta$ corresponds to its normal component. Both components show a good correlation with $F_\mathrm{w}$. The correlation coefficient $r$ is greater than 0.88 in both cases. The raw $F_\mathrm{t}$ shows good correlation with $\langle F_\mathrm{w}\rangle_{300\mbox{--}500}$ ($r=0.85$). However, we can find the best correlation in $F_\mathrm{t} \sin\theta$ vs. $F_\mathrm{w}$ (Fig.~\ref{fig:correlations}(a)]. Thus, the shear component of the force chain plays the most important role in resisting the rod withdrawal.

In a previous work~\cite{Furuta:2019}, Furuta et al. proposed a model of an effectively solidified zone that can be constructed by obliquely extending force chains. In this experiment, we directly observed the development of oblique force chains. However, clear evidence of solidification cannot be obtained solely from photoelastic information. To quantitatively discuss the mechanism of the withdrawal force increase, it seems that a much larger experimental system has to be investigated. 

\section{Conclusion}
In this study, the development of a force-chain structure in a 2D granular layer, from which a rod was vertically withdrawn, was experimentally studied using photoelastic disks. By only using the force chains developed by the rod withdrawing, the characteristic quantities of force chains were measured. Specifically, the average total force and angle of the force chains were computed from photoelastic images. With the rod withdrawing, force chains with a roughly constant inclination angle $\theta \simeq 20^{\circ}$ were rapidly developed, and the total force of the force chains gradually increased by the rod withdrawal. To characterize the relationship between rod withdrawal and force-chain development, the data in the approximately steady withdrawing regime, $300 \leq t \leq 500$~s, were analyzed in detail. In the relatively large initial packing fraction regime ($\phi_0 > 0.825$), we confirmed a $\phi_0$-dependent increase in the average total force $F_\mathrm{t}$ and rod-withdrawing force $F_\mathrm{w}$. Moreover, a correlation between $F_\mathrm{t}$ and $F_\mathrm{w}$ was clearly observed in this regime. In particular, the shear component $F_\mathrm{t}\sin\theta$ showed the best correlation with $F_\mathrm{w}$. The obtained result is consistent with that of a previous study~\cite{Furuta:2019}. However, more quantitative analysis with a much larger experimental system is necessary to discuss the details of the force-chain development in the rod-withdrawn granular layer.

\section*{Acknowledgments}

We thank JSPS KAKENHI Grant No.~18H03679 for financial support.

\bibliographystyle{ws-mplb}
\bibliography{pew}

\begin{thebibliography}{10}
\newcommand{\enquote}[1]{#1}

\bibitem{Behringer:2014}
R.~P. Behringer, D.~Bi, B.~Chakraborty, A.~Clark, J.~Dijksman, J.~Ren and
  J.~Zhang, \emph{J. Stat. Mech.} \textbf{2014} (2014) P06004.

\bibitem{Bi:2011}
D.~Bi, J.~Zhang, B.~Chakraborty and R.~P. Behringer, \emph{Nature} \textbf{480}
  (2011) 355.

\bibitem{Majmudar:2005}
T.~S. Majmudar and R.~P. Behringer, \emph{Nature} \textbf{435} (2005) 1079.

\bibitem{Ren:2013}
J.~Ren, J.~A. Dijksman and R.~P. Behringer, \emph{Phys. Rev. Lett.}
  \textbf{110} (2013) 018302.

\bibitem{Wang:2018}
D.~Wang, J.~Ren, J.~A. Dijksman, H.~Zheng and R.~P. Behringer, \emph{Phys. Rev.
  Lett.} \textbf{120} (2018) 208004.

\bibitem{Oda:1974}
M.~Oda and J.~Konishi, \emph{Soils and Foundations} \textbf{14} (1974) 25.

\bibitem{Marone1998}
C.~Marone, \emph{Annu. Rev. Earth Planet. Sci.} \textbf{26} (1998) 643.

\bibitem{GDRMiDi:2004}
{GDR~MiDi}, \emph{Eur. Phys. J. E} \textbf{14} (2004) 341.

\bibitem{daCruz2005}
F.~da~Cruz, S.~Emam, M.~Prochnow, J.-N. Roux and F.~Chevoir, \emph{Phys. Rev.
  E} \textbf{72} (2005) 021309.

\bibitem{Jop2006}
P.~Jop, Y.~Forterre and O.~Pouliquen, \emph{Nature} \textbf{441} (2006) 727.

\bibitem{Pouliquen2006}
O.~Pouliquen, C.~Cassar, P.~Jop, Y.~Forterre and M.~Nicolas, \emph{J. Stat.
  Mech.}  (2006) P07020.

\bibitem{Kuwano2013}
O.~Kuwano, R.~Ando and T.~Hatano, \emph{Geophys. Res. Lett.} \textbf{40} (2013)
  1295.

\bibitem{Furuta:2017}
T.~Furuta, K.~Katou, S.~Itoh, K.~Tachibana, S.~Ishikawa and H.~Katsuragi,
  \emph{Int. J. Mod. Phys. B} \textbf{31} (2017) 1742006.

\bibitem{Furuta:2019}
T.~Furuta, S.~Kumar, K.~A. Reddy, H.~Niiya and H.~Katsuragi, \emph{New J.
  Phys.} \textbf{21} (2019) 023001.

\bibitem{Howell:1999}
D.~Howell, R.~P. Behringer and C.~Veje, \emph{Phys. Rev. Lett.} \textbf{82}
  (1999) 5241.

\bibitem{Iikawa:2015}
N.~Iikawa, M.~M. Bandi and H.~Katsuragi, \emph{J. Phys. Soc. Jpn.} \textbf{84}
  (2015) 094401.

\bibitem{Iikawa:2018}
N.~Iikawa, M.~M. Bandi and H.~Katsuragi, \emph{Phys. Rev. E} \textbf{97} (2018)
  032901:1.

\bibitem{Jotaki:1977}
T.~Jotaki and R.~Moriyama, \emph{J. Soc. Powder Tech. Jpn.} \textbf{14} (1977)
  609 , in Japanese.

\bibitem{Bertho:2003}
Y.~Bertho, F.~Giorgiutti-Dauphin\'e and J.-P. Hulin, \emph{Phys. Rev. Lett.}
  \textbf{90} (2003) 144301.

\bibitem{Katsuragi:2016}
H.~Katsuragi, \emph{Physics of Soft Impact and Cratering} (Springer, 2016),
  {Chapter 3.7}.

\end{thebibliography}

\end{document}